\documentclass[amsmath,amssymb, aps,twocolumn, superscriptaddress]{revtex4-2}

\usepackage{float}
\usepackage{graphicx}
\usepackage{dcolumn}
\usepackage{bm}
\usepackage{gensymb}
\usepackage[normalem]{ulem}

\usepackage{textcomp}
\usepackage{mathptmx}
\usepackage{amsmath}
\usepackage{appendix}
\usepackage{color}

\usepackage[dvipsnames]{xcolor}

\usepackage{diagbox}
\def\theTitle{An optically pumped magnetic gradiometer for the detection of human biomagnetism}

\def\PA{School of Physics and Astronomy, University of Birmingham, \\ Edgbaston, Birmingham, B15 2TT, United Kingdom}
\def\CHBH{Centre for Human Brain Health,  School of Psychology, University of Birmingham, \\ Edgbaston, Birmingham, B15 2SA, United Kingdom}

\begin{document}

\title{\theTitle} 
\author{Harry Cook}
\affiliation{\PA}
\author{Yulia Bezsudnova}
\affiliation{\PA}
\author{Lari M. Koponen}
\affiliation{\CHBH}
\author{Ole Jensen}
\affiliation{\CHBH}
\author{Giovanni Barontini}
\affiliation{\PA}
\affiliation{\CHBH}
\author{Anna U. Kowalczyk}
 \email{a.u.kowalczyk@bham.ac.uk}
\affiliation{\CHBH}
\date{\today}

\begin{abstract}
We realise an intrinsic optically pumped magnetic gradiometer based on non-linear magneto-optical rotation.  We show that our sensor can reach a gradiometric sensitivity of 18~fT$/\text{cm}/\sqrt{\text{Hz}}$ and can reject common mode homogeneous magnetic field noise with up to 30~dB attenuation. We demonstrate that our magnetic field gradiometer is sufficiently sensitive and resilient to be employed in biomagnetic applications. In particular, we are able to record the auditory evoked response of the human brain, and to perform real-time magnetocardiography in the presence of external magnetic field disturbances. Our gradiometer provides complementary capabilities in human biomagnetic sensing to optically pumped magnetometers, and opens new avenues in the detection of human biomagnetism. 
 
\end{abstract}

\maketitle

\section{Introduction} \label{sec:Intro}
Highly sensitive magnetometers \citep{Grosz16} have a diverse range of applications that span from geophysics and exploration \citep{Love08}, to non-destructive magnetic materials testing \citep{GM11,Romalis11,Bevington18}, archaeology and palaeomagnetism \citep{Fassbinder2017}, environmental monitoring \citep{Fu20}, navigation and positioning \citep{Li14}, space exploration \citep{Bennett21}, biology, neuroscience  \citep{Budker2013, Tierney19, Aslam23}, and fundamental physics research \citep{Budker2013, Kimball23}. Over the past few decades, advances in quantum science spurred the development of various types of magnetometers with very different operating principles: from electron spin resonance magnetometers such as SQUIDS to optically pumped magnetometers with NV centers in diamond or thermal atoms. 
 
Optically pumped magnetometers (OPMs)  based on thermal atoms are able to achieve subfemtotesla sensitivities \citep{Kominis03} and do not require cryogenics, so that they are compact, portable and inexpensive. Therefore, they are emerging as the preferred sensors in biomagnetism applications previously dominated by SQUIDs \citep{Xia2006, Boto2016, Broser18,Jensen2018,Bu22,Xiao2023}. In human brain imaging in particular, new capabilities have been demonstrated when OPMs are used for magnetoencephalography (MEG) \cite{IIVANAINEN2019,Labyt2019,Boto2022,Alem23}. MEG normally requires the detection of both magnetic fields and magnetic field gradients, with these latter enabling higher spatial resolution and better localization of the biomagnetic source \citep{VRBA2002, Koga13, Marhl22, Wens23}. In general, the measurement of the small magnetic fields produced by the human body is strongly affected by environmental magnetic field noise, whose cancellation and shielding is expensive and cumbersome \cite{IIVANAINEN2019,Holmes19, Jazbinsek22,Holmes2023}. Measuring magnetic field gradients can help in circumventing or simplifying this problem. 

Several types of optically pumped magnetic gradiometer (OPMG) have been realised. In the so-called synthetic gradiometers, signals from two or more closely spaced sensors are subtracted digitally or electronically. Such a configuration suppresses the common mode magnetic field noise, but enables only reduced sensitivity \citep{Zhang16}. Semi-intrinsic OPMGs use a single laser beam split into either two \citep{Sheng17} or multiple measurement channels \citep{Kim14}. The signals produced are electronically subtracted enabling superior sensitivity. Intrinsic OPMGs directly subtract common-mode magnetic field noise before converting the magneto-optical rotation signal to a photocurrent. This eliminates the need of post-processing and does not degrade the sensor's sensitivity \citep{Wasilewski10, Kamada15,Perry2020, Zhang20,Lucivero21, Cooper22}. 

In this work, we realise an intrinsic OPMG based on non-linear magneto-optical rotation (NMOR). We show that our OPMG can reach sensitivities of 18~fT$/\text{cm}/\sqrt{\text{Hz}}$ while being resilient to fluctuations of external magnetic fields, that can be suppressed up to 30 dB. We demonstrate that such an OPMG is sufficiently sensitive and resilient to be employed in human biomagnetic applications such as MEG and magnetocardiography. In particular, we are able to measure the magnetic field gradient produced by the auditory evoked field in the human brain, and the magnetic field gradient produced by cardiac activity in the presence of external magnetic field disturbance. 

This work is organised as follows: in section \ref{sec:theory} we describe the working principle of our OPMG, in section \ref{sec:Design} we detail the design of our sensor, in section \ref{sec:performance} we report the testing of the performance of our sensor in a controlled environment, in section \ref{sec:biosensing} we present the applications of our sensor in MEG and magnetocardiography experiments, and in in section VI we report our conclusions.

\section{Working principle} \label{sec:theory}

In an NMOR magnetometer, amplitude or frequency-modulated linearly polarised resonant light is used to induce spin precession in an atomic gas. A static, homogeneous bias magnetic field applied along the direction of propagation of the light sets the Larmor frequency. Changes in the external magnetic field increase or decrease the Larmor frequency, and are detected by monitoring the rotation of the light polarisation that happens synchronously with the modulation \citep{Pustelny2008}. 

\begin{figure}[t]
\includegraphics[width=\columnwidth]{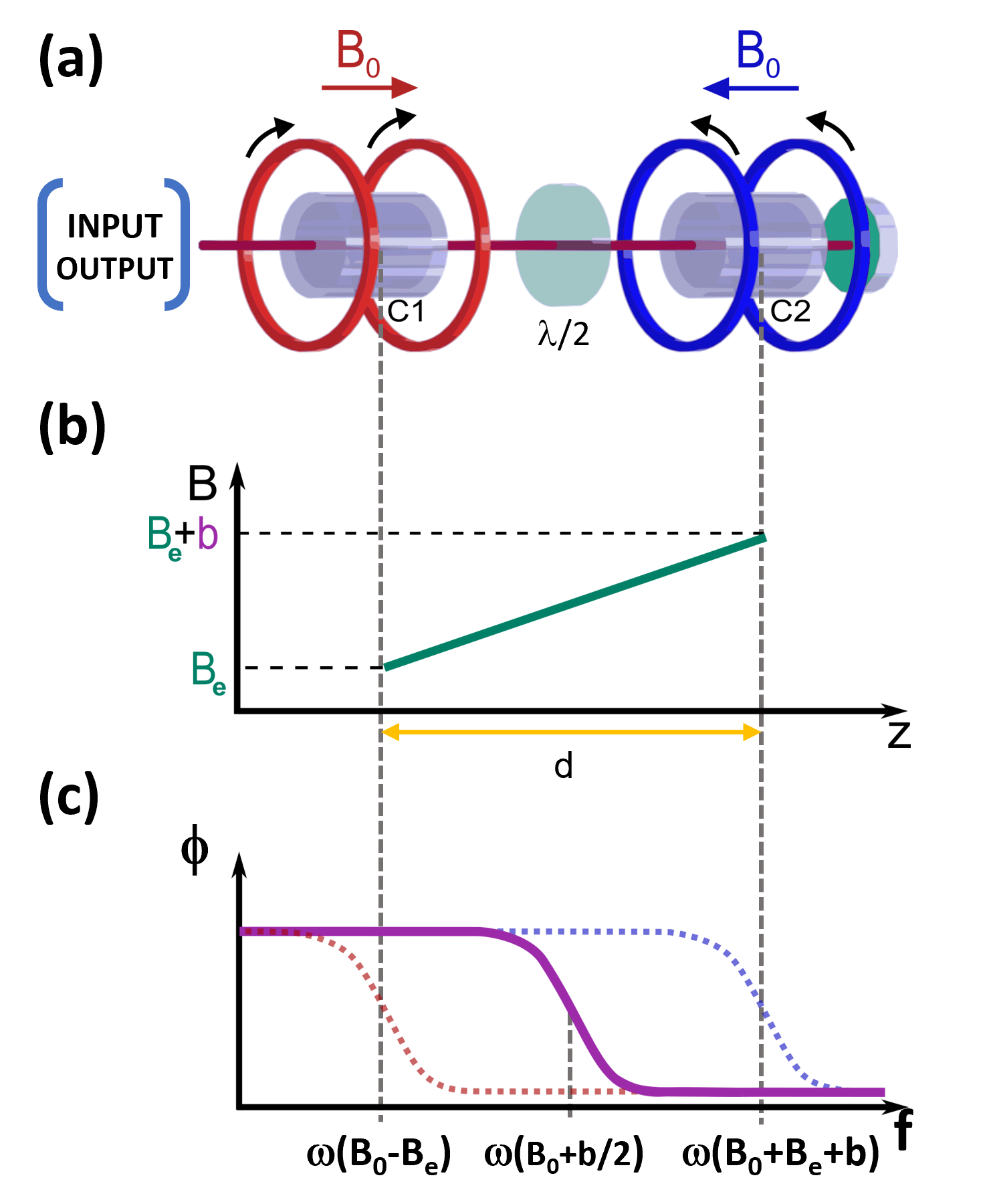}
 \caption{\label{fig:overlapping} Intrinsic gradiometer operating principle. (a) A laser beam passes through two cells (C1 and C2) with equal bias fields $\vec{B}_{0}$ pointing in opposite directions and is retroreflected on a mirror. The half-waveplate ($\lambda$/2) between the cells flips the sign of the resonance generated in C2 allowing constructive addition of NMOR resonances. (b) Magnetic fields along the sensitive axis of the gradiometer: $B_e$ is the homogeneous external field and $b$ is the differential magnetic field between two cells separated by the gradiometer baseline $d$. (c) The continuous line is the total phase of the gradiometer in (a) as a function of the frequency, in the presence of the magnetic field in (b). The dotted lines show the behaviour of the phase in the case the cells were probed independently. }
\end{figure}

The working principle of our OPMG is an extension of the NMOR technique. Fig. \ref{fig:overlapping}(a) shows the basic elements of the gradiometer: a near-resonant laser beam passes through two vapour cells and is retroreflected. The bias fields in the two cells have the same magnitude $B_0$ but opposite directions. A waveplate between the two cells rotates the polarisation by $\pi$, ensuring constructive addition of the NMOR resonances. Supposing that, as shown in Fig. 1 (b), the gradiometer is immersed in an external magnetic field with amplitude $B=B_e+B'z$ directed along the direction of propagation of the light $z$, the signal produced is the sum of the NMOR signals in each cell:
\begin{equation}
    Q_{G}+iP_{G}= [Q_{C1}+Q_{C2}]+i[P_{C1}+P_{C2}]
\end{equation}
where $Q_G$ and $P_G$ are the total quadrature and in-phase signals, and $Q_{Ci}$ and $P_{Ci}$ denote the quadrature and in-phase components of the signal produced in each cell. By writing the two contributions explicitly, we obtain:
\begin{eqnarray}
    Q_G(\omega) &=& A_1\left(\frac{\Gamma_1}{2}\right)^{2}\frac{1}{[\omega-\omega(B_0+B_e+b)]^{2} + (\Gamma_1/2)^{2}} \nonumber \\
    &+&A_2\left(\frac{\Gamma_2}{2}\right)^{2} \frac{1}{[\omega-\omega(B_0-B_e)]^{2} + \left(\Gamma_2/2\right)^{2}}
\end{eqnarray}

\begin{eqnarray}
    P_G(\omega) &=& A_1 \frac{\Gamma_1}{2} \frac{\omega-\omega(B_0+B_e+b)}{[\omega-\omega(B_0+B_e+b)]^{2} + (\Gamma_1/2)^{2}} \nonumber \\ 
    &+&A_2\frac{\Gamma_2}{2} \frac{\omega-\omega(B_0+B_e)}{[\omega-\omega(B_0-B_e)]^{2} + \left(\Gamma_2/2\right)^{2}},
\end{eqnarray}
where $A_i$ and $\Gamma_i$ are the amplitude and width of the NMOR resonance in cell $i$, $\omega(B)=g\mu_BB/\hbar$ are the NMOR resonance frequencies, $\mu_B$ the Bohr magneton, $g$ the Land\'e g-factor, $\hbar$ the reduced Planck constant, and $b=B'd$, with $d$ the separation between the cells. Assuming that $A_1=A_2$ and $\Gamma_1=\Gamma_2>\omega(b+2B_e)$, we find the zero of the total phase
\begin{equation}
    \Phi(\omega) =\textrm{arctan}\left(\frac{P_G}{Q_G}\right) = \frac{P_G}{Q_G}
\end{equation}
to be at
\begin{equation} \label{eq:half-omegab}
    \omega = \omega(B_0 +b/2).
\end{equation}
The position of the gradiometric NMOR resonance is therefore independent on $B_e$, and depends only on the differential magnetic field between the cells $b$ (Fig. 1 (c)). If the separation between the cells $d$ is known, the gradiometer provides a direct measurement of $B'$.

\section{The NMOR gradiometer sensor} \label{sec:Design}

\begin{figure}[t]
\includegraphics[width=\columnwidth]{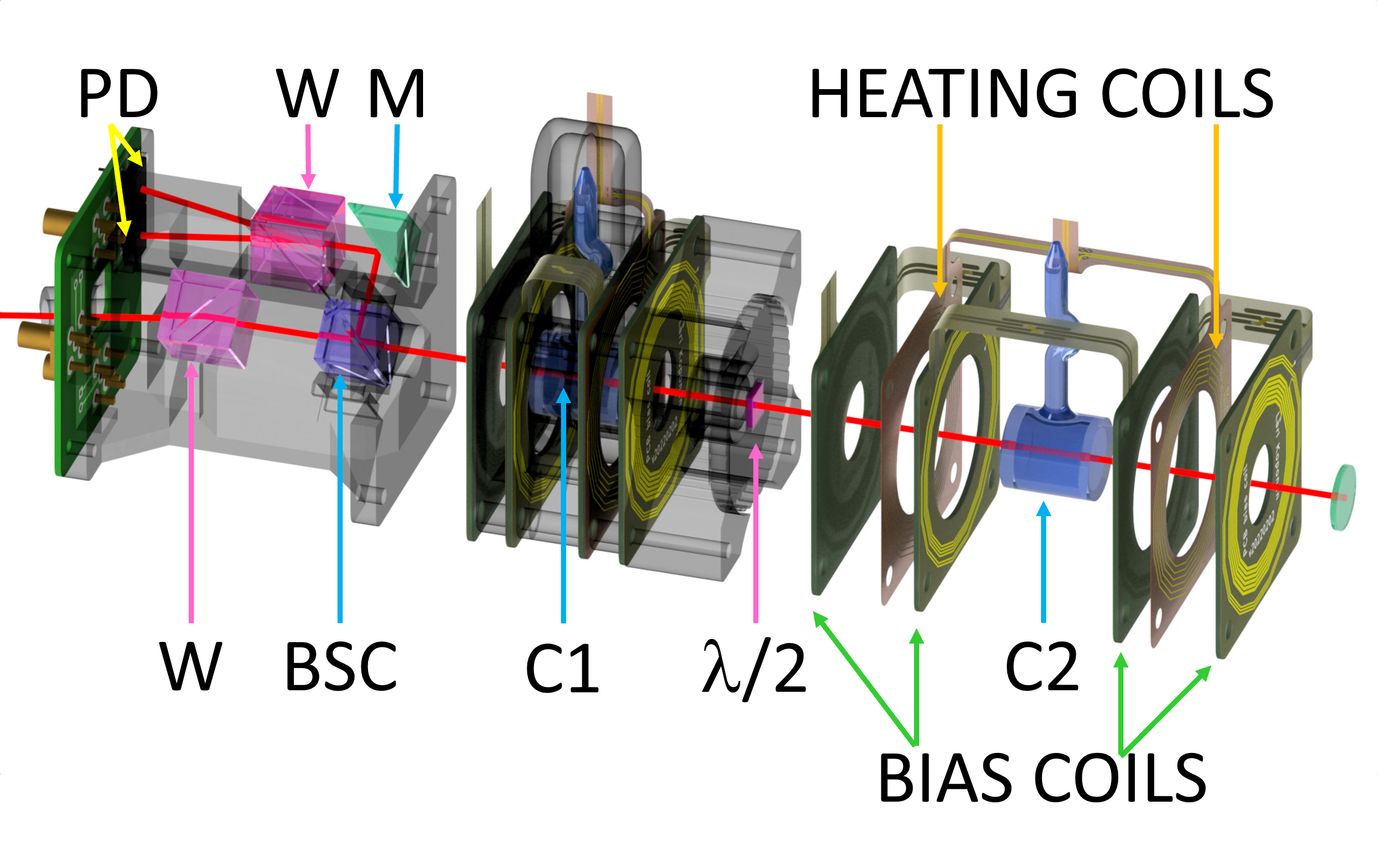}
\caption{\label{fig:CAD}Sensor layout: exploded view. The laser beam (represented with a red line) is delivered with an optical fibre and collimated using a GRIN lens (not shown). The beam is deflected on a Wollaston prism (W), 10\% of the beam passes through the beam splitting cube (BSC), cell 1 (C1), half-wave plate  ($\lambda/2$),  cell 2 (C2)  and is retroreflected with a mirror (M). After passing through the cells the second time, 90\% of the beam is reflected by the BSC and then a right-angle mirror to the polarimeter formed by a second Wollaston prism and two photodiodes (PD). The coils surrounding both cells are identical but are expanded around C2 for better visualisation. }
\label{fig:sensor-design} 
\end{figure}

The sensor head of our OPMG is an evolution of the NMOR magnetometer described in \cite{Kowalczyk2021}. All the components of the sensor and the optical path are shown in Fig. \ref{fig:sensor-design}.   
 The laser light is delivered by a polarisation-maintaining optical fibre, and is collimated to a 1.8~mm beam diameter using a non-magnetic GRIN (graded-index) lens. The polarisation of the beam is cleaned with a Wollaston prism (W) that also aligns the beam along the sensor axis. The beam passes through a 10/90 non-polarizing beam splitter cube (BSC), and then through the two sensing cells, that are separated by a half-waveplate ($\lambda$/2). The beam is retro-reflected by a plane mirror and, after passing through the cells a second time, is reflected by the BSC to the polarimeter. This consists of another Wollaston prism, rotated by 45$^{\circ}$, and a balanced photodiode (PD). The PD signal is delivered to a transimpedance amplifier, which is then fed to a lock-in amplifier. To track the changes of the measured magnetic field in real-time, we use a combination of lock-in detection and a phase-locked loop (PLL) that ensures that the resonance condition is always fulfilled. 
 
The two sensing cells (C1 and C2) are paraffin-coated and have cylindrical internal dimensions of 1.00~cm in length and 0.95~cm in diameter \citep{BEZSUDNOVA22}. The centres of the two cells are separated by $d$~=~4~cm. The baseline was chosen as a compromise between measuring magnetic field gradients originating from the brain and keeping the sensor compact. Each cell is held in place with a polyjet 3D-printed mount made of thermoset acrylic resin. To produce the bias field in each cell we use a pair of \emph{self-shielded} bias coils, that generate reduced external stray fields and are also least sensitive to external electromagnetic noise due to reciprocity. The first feature is especially important in the case of gradiometers, which have two adjacent cells with bias fields that can affect each other. Each bias coil is made from a rigidised flexible PCB, consisting of 4 elements, placed symmetrically around each cell (Fig. 2). Each element has two layers of copper. The spacing between the elements and the geometry within each layer are optimised numerically for the best possible balance between field homogeneity and self-shielding factor. Compared to a solenoid of the same diameter and length, the optimised bias coils produce a more homogeneous field (mean magnetic field inhomogeneity over the cell volume 1.2\% versus 6.0\% for our earlier solenoid) and reduce the stray field by 20~dB at 4.0~cm (i.e., at the other cell of the gradiometer or for two adjacent sensors) and by $>$ 40~dB at $>$ 8~cm. Consequently, the mean magnetic field of bias coil 1 on cell C2 is reduced from $-11\%$ to $-1.2\%$ and the mean magnetic field inhomogeneity in the gradiometer configuration is improved to 1.8\% (versus 8.2\% for a pair of our earlier solenoids).

Because of the light absorption within the cells, the power of the laser beam decreases after each pass. Since the amplitude of the NMOR signals depends on the light intensity, this results in $A_1\neq A_2$, voiding a necessary condition for the correct functioning of the gradiometer. To equalise the amplitude of the individual NMOR signals, we independently control the optical density of the atomic vapour in each cell with the temperature. This is achieved with ac current heating and dedicated thin two-layer flexible PCB coils (Fig. 2). The winding pattern for the heating coils is optimised to produce the lowest possible magnetic field at each cell (i.e., a bifilar coil with winding number zero for both filaments). Coincidentally, a coil in such configuration produces internal stray fields approximately orthogonal to the sensitive axis of the gradiometer. In NMOR, fields normal to the bias field and of smaller amplitude do not significantly affect the sensitivity \citep{Pustelny2006}. Additionally, we drive the coils with low-noise audio amplifiers at 21~kHz, which is far detuned from the NMOR resonance at $\sim$1.5~kHz. The temperature in each cell is measured with a non-magnetic PT1000 resistance temperature detector, which is connected to a low-noise readout amplifier. This readout is delivered to a microcontroller which outputs a digital signal of the measured temperature to a PC. The amplitude of the two 21~kHz sinuoids is stabilized with a digital feedback loop. The signals are passed through an audio amplifier to generate the current needed to heat the cells. Due to the proximity of the temperature sensor to the cells, probing the temperature generates spurious magnetic field noise, therefore we sample the temperature only every 15~s. This allows us to control the temperature with 0.1~$^{\circ}$C precision. 

\section{Performance} \label{sec:performance}

\begin{figure}[t]
\includegraphics[width=\columnwidth]{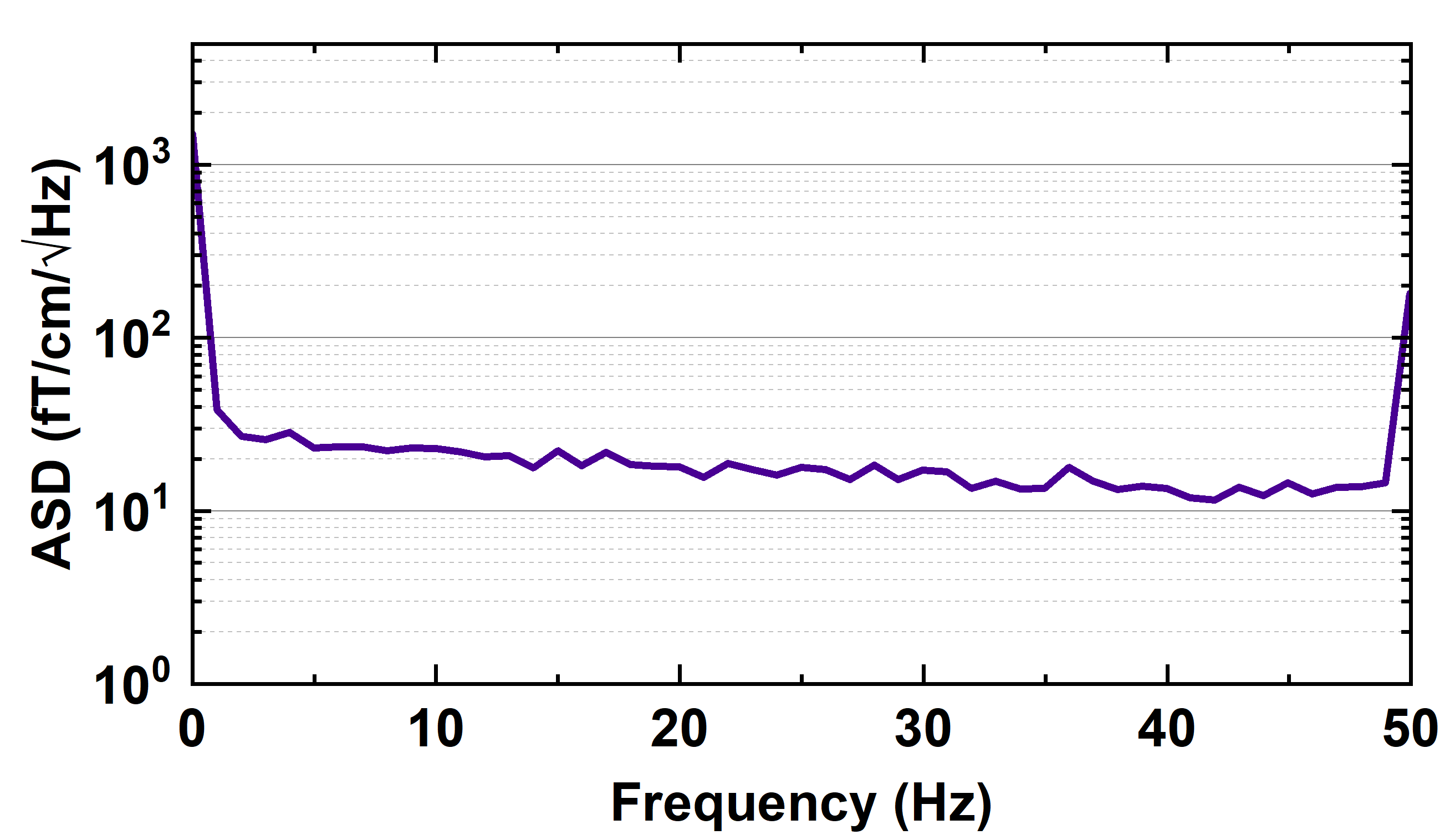}
\caption{\label{fig:sens} Amplitude spectral density (ASD) of our sensor as a function of frequency. The data shown are averaged over 30 one-second experimental traces. }
\end{figure}

\begin{figure*}[ht!]
    \centering
    \includegraphics[width=\textwidth]{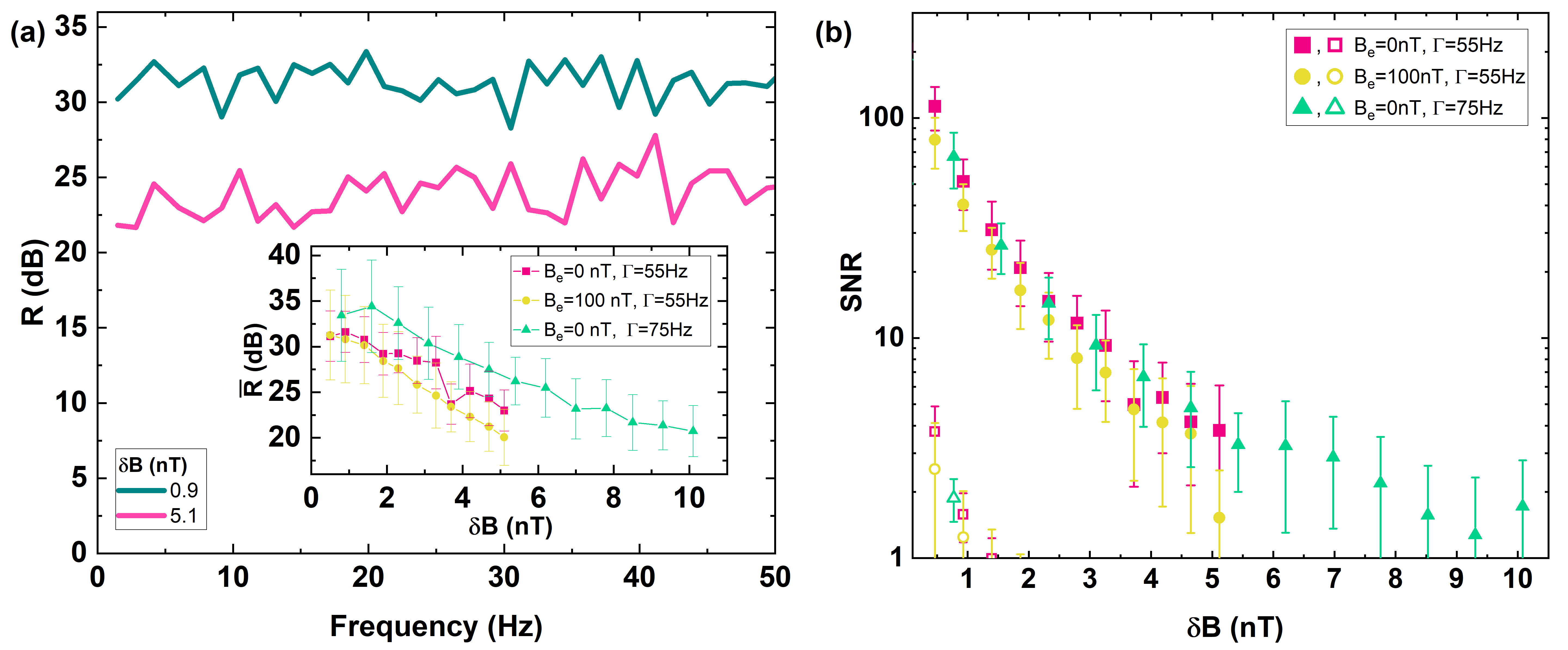}
    \caption{\label{fig:Rejection} (a) Rejection coefficient $R$ of our sensor as a function of the frequency, for two different amplitudes of the magnetic field noise $\delta$B, measured in a controlled environment. The amplitude of the external magnetic field for this measurement was set to 0.9~nT and 5.1~nT, the set width of the gradiometer signal was 55~Hz. The inset shows the averaged rejection coefficient $\overline{R}$ over 1-20~Hz as a function of the amplitude of the applied magnetic field noise in experimental conditions (i) $B_e$ =0~nT and $\Gamma$=55~Hz (pink squares), (ii) $B_e$ =100~nT and $\Gamma$=55~Hz (yellow dots), (iii) $B_e$ =0~nT and $\Gamma$=75~Hz (green triangles). (b) Signal-to-noise ratio of a detected sinusoidal gradient field oscillating at 6 Hz for the gradiometer (filled points) and magnetometer (empty points) as a function of the amplitude of the applied magnetic field noise. }
    \label{fig:atten}
\end{figure*}

To characterise the performance of our sensor, we placed it at the centre of a cylindrical 4-layer $\mu$-metal shield that has an internal coil system for precise control of the magnetic fields (Twinleaf MS-2). To obtain equal resonance amplitudes, we first warm up C1 to 42 $^{\circ}$C and C2 to 38 $^{\circ}$C, and then fine-tune the temperatures until the amplitudes are equal within 5$\%$ error. Finally, we overlap the resonances at a single frequency in the range of 1.5-2~kHz by adjusting the bias fields in each cell. In this work,  the lock-in detection is performed using a third-order digital low-pass filter with a time constant of 1.61 ms. The signal acquired from the PLL is sampled with a rate of 837.1~Hz. 

In line with convention, we define the sensitivity of our sensor as the measured average noise floor of the amplitude spectral density. Specifically we measure the noise floor in the 2-48 Hz band, which is the band relevant for biomagnetic measurements. We determine the experimental sensitivity of our sensor by recording 30 one-second traces in the absence of any applied magnetic fields. We convert the signal time-course to amplitude spectral density for each trace and then average these traces. The result is shown in Fig. \ref{fig:sens}. We measured a sensitivity of (17.9~$\pm$~1.4)~fT$/\text{cm}/\sqrt{\text{Hz}}$, which is mainly limited by the electronic and PLL noise. 

Another important parameter of the sensor is its insensitivity to temporal variations of homogeneous magnetic fields. To determine the resilience of our sensor, we measured its common mode rejection coefficient $R=20\textrm{log}_{10}\left[ASD_g(\omega)/ASD_m(\omega)\right]$, with $ASD$ the amplitude spectral density, and with the subscripts $g$ and $m$ indicating the sensor operating in gradiometer or magnetometer mode respectively \footnote{Operating the sensor in magnetometer mode was done by switching off the bias field of C1}. To measure $R$, we applied a homogeneous magnetic white noise of amplitude $\delta B$ along the sensor's sensitive direction. This was produced with a low-noise voltage-driven current source controlled by an arbitrary waveform generator. The applied white magnetic field noise had a bandwidth of 1~-~100~Hz, and $\delta$B was varied in the 0.5~-~10.1~nT range. For each $\delta$B, we recorded 10 traces of 3 s, and repeated the measurements for 3 sets of operating conditions:
\begin{itemize}
    \item Setup (i): we tuned the laser power and detuning to achieve the best possible sensitivity. The resulting minimum full width at half maximum of the gradiometer resonance $\Gamma$ was (55~$\pm$~3)~Hz. 
    \item Setup (ii): to investigate the resilience of our sensor to high external magnetic fields, we additionally applied a DC offset of $B_e$~=~100~nT along the sensitive direction. This procedure did not affect $\Gamma$ which remained $\sim$55~Hz. 
    \item Setup (iii): to explore the relation between bandwidth and sensitivity of NMOR and light intensity, $\Gamma$ was increased to 75~$\pm$~3~Hz.
\end{itemize}
For each trace, we calculated the amplitude spectral density and then we performed the average over 10 traces. These averages were then used to evaluate $R$. We also calculated the averaged rejection $\bm{\overline{R}}$ by averaging $R$ across the 1~-~20~Hz band. 

Fig. \ref{fig:Rejection} summarises the performance of the OPMG in all 3 operating conditions. In panel (a) we show $R$ for $\delta B$~=~0.9~nT (pink line) and $\delta B$~=~5.1~nT (green line), obtained in the best sensitivity condition (i). For each $\delta B$, $R$ remains fairly constant over the entire frequency range: $\bm{\overline{R}}$ is $\sim$~31~dB for $\delta$B~=~0.9~nT and $\sim$~23~dB for $\delta$B and 5.1~nT.  In the inset of Fig. \ref{fig:Rejection} (a), we show the dependence of $\bm{\overline{R}}$ on $\delta B$ for the 3 operating settings. The pink squares are data collected in settings (i). The attenuation for $\delta B>$ 1.5~nT linearly decreases. This is because the overlap between the C1 and C2 resonances is reduced by $\sim$ 30\%, and therefore the condition $\Gamma_1=\Gamma_2>\omega(b+2B_e)$ is no more well verified. Beyond $\delta B~\simeq$~1.5~nT, such a condition is no more satisfied and the noise rejection of the sensor rapidly degrades. The absolute maximum $\delta B$ that our gradiometer can tolerate in settings (i) is $\sim$~5~nT.  The yellow circles represent the data in settings (ii). Since increasing the background field does not significantly affect $\Gamma$, the performance of the gradiometer is similar to (i), with slightly worse $\bm{\overline{R}}$ at higher $\delta$B. This demonstrates the ability of our sensor to operate in high external magnetic fields, a characteristic `inherited' by the underlying NMOR physics. The green triangles are the data in settings (iii). Here the rejection range is increased up to  $\delta B~\simeq$~10.1 nT, and $\bm{\overline{R}}$ is increased above 30 dB for $\delta$B~$<$~2.5~nT. This is a result of wider $\Gamma$, so that the condition $\Gamma_1=\Gamma_2>\omega(b+2B_e)$ is stringently verified over a broader range.

To determine the ability of the sensor to measure a magnetic field gradient in the presence of external magnetic field noise, we repeated the above measurements additionally applying  a small sinusoidal magnetic field gradient $B'$~=~5~pT$/$cm, oscillating at 6~Hz. To extract the signal-to-noise ratio (SNR) we divide the amplitude of the measured signal at 6~Hz with the measured average noise in the 10~-~50~Hz range, and then average over 10 repetitions. The results are reported in Fig. \ref{fig:Rejection} (b). The solid points are the data collected in the gradiometer mode, while the open points in magnetometer mode. For all the settings investigated, the SNR computed for the OPMG outperforms the one of the OPM by more than one order of magnitude for $\delta B~<~1$~nT. For higher values of $\delta B$, it is not possible to identify a measurable signal at 6~Hz when operating in magnetometer mode. 

\section{Biomagnetic sensing} \label{sec:biosensing}
Measurement of human biomagnetism is very demanding as it requires both high sensitivity and substantial shielding from external perturbations. These requirements are often met by employing a combination of highly sensitive magnetic gradiometers inside a magnetically shielded room. Our OPMG has the potential to intrinsically satisfy both requirements because, as we show in the following, it is sufficiently sensitive to detect human brain activity, and sufficiently resilient to detect human cardiac activity in a substantially perturbed environment.

\subsection{Magnetoencephalography} \label{MEG}

MEG is a non-invasive neuroimaging technique that measures weak magnetic fields produced by neural activity in the brain \citep{Hamalainen1993}. To test the response of our gradiometer to brain signals we measured event-related fields (ERF) in the human brain. The experiment took place at the University of Birmingham, Centre for Human Brain Health. The participant was seated inside a  2-layer magnetically shielded room. We recorded the auditory evoked field (AEF) to a binaural oddball paradigm \citep{Garrido2008} commonly used in benchmarking OPM sensors \citep{Borna2017, Seymour2021,Iivanainen2023}. The research protocol was approved by the Science, Technology, Engineering and Mathematics Ethical Review Committee at the University of Birmingham. The participant was informed about the experimental method and a written consent was obtained. 

The participant was seated in a custom-made wooden chair with an adjustable platform to attach the sensor. The auditory stimulus produces a strong response in the brain, with a well-defined activity region making it easy to position the sensor around the head. Similarly to \citep{Kowalczyk2021}, we have first determined the location on the subject's scalp with the highest AEF using a conventional MEG system. The gradiometer was then positioned at this location. During both sessions, the participant was listening to two auditory stimuli. A 1~kHz pure tone was played for 80\% of the trials, whilst during the remaining 20\%, the oddball 40 Hz thump was played. The oddball tone not only induces the AEF itself, but it also increases the brain response from the subsequent normal 1 kHz tone. The duration of the stimuli was 100 ms followed by an interval randomly varied between 911 ms and 1111 ms. This randomised time gap between stimuli was introduced to prevent the adaptation of the brain to the repeated noise. The sound was generated using a SOUNDPixx system and was delivered to the participant binaurally using MEG-compatible air-tubes and disposable ear-pieces. No active response from the participant was required. The participant was asked to concentrate on the tone, fixate eyes on the target placed on the wall and minimise any movement. We recorded 445 trials with 837 Hz sampling rate. Along with the gradiometer signal, we recorded an analogue trigger signal to timestamp the presentation of the tone. 

\begin{figure}[t]
\includegraphics[width=\columnwidth]{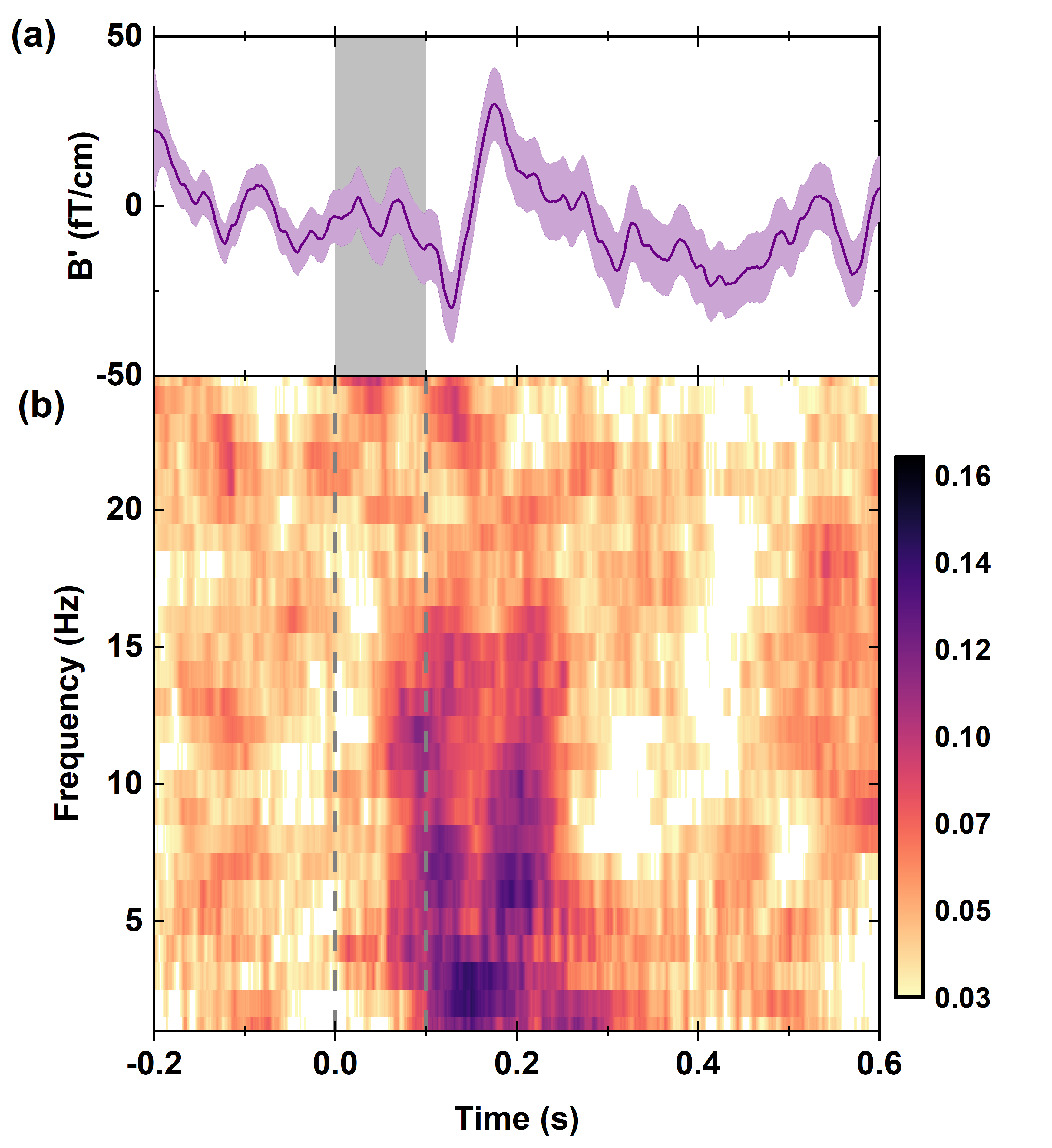} 
\caption{\label{fig:brain} Time course of the event-related field (a) The purple line is the Auditory Evoked Response averaged over 445 epochs, the shading represents the standard error of the mean. (b) Inter-trial coherence across all epochs in the time-frequency domain. The gray shaded area in (a) and the vertical grey lines on (b) show the onset and duration of the auditory tone.}
\end{figure}

The raw data were detrended and epoched. A pre-tone interval of 200~ms was used to perform baseline correction, which subtracted all values within an epoch by the average baseline value, correcting for dc offsets. The line noise was removed by fitting and subtracting a single sinusoidal component near 50~Hz from each epoch. All epochs were filtered with a 20~Hz first-order Butterworth infinite impulse response  filter (operating in both forward and reverse directions to achieve an acausal response with zero phase shift) and averaged. 

The typical time course of the AEF obtained with our gradiometer is shown in Fig. \ref{fig:brain} (a). The purple line shows the averaged signal while the purple shaded area represents the standard error of the mean calculated across all epochs. The grey shaded area indicates the stimulation duration. At about 100~ms a strong deflection can be noticed that corresponds to the N100m peak \cite{Hamalainen1993}, followed by another one at 150ms corresponding to P150m. The total amplitude of the detected N100 and P150 peaks is 60~fT/cm , yielding an averaged signal-to-noise ratio of 5.7 within this region. The inter-trial coherence (ITC) was calculated for all AEF epochs and the results are shown in Fig.~\ref{fig:brain} (b). ITC in MEG signals provides information about the consistency or phase-locking of neural activity across different trials of an experiment. 
ITC is acquired here by applying time-frequency decomposition on each trial with a single taper applied to each 250~ms long sliding window. The higher coherence values represent a correlation in the phase of the signal across epochs. The spectral leakage into neighbouring frequency intervals is a side-effect of the tapering method and windowing parameters used in order to maintain good temporal resolution. Nonetheless, there is a clear increase in ITC during the same 100~-~200~ms period after stimulus onset which confirms consistent synchronisation in the activity of neural ensembles in response to auditory stimuli during the task. 

\subsection{Magnetocardiography} \label{MCG}

To test the performance of our sensor in real-time recordings, and to demonstrate the attenuation of environmental magnetic noise, we measured the heartbeat of a human participant during a time in the day when there was intense activity in the building. Our magnetically shielded room is located in the basement of a 7-story building in the vicinity of lifts that create up to 5~nT peak-to-peak magnetic field changes when running the full distance. In this experiment, the gradiometer was fixed on the platform of the wooden chair while the participant was leaning over the sensor keeping a 5~cm distance from the sensor surface. The data acquisition was carried out with the same lock-in parameters as in the MEG session, however, no filtering or averaging has been applied. During recording, the building lifts were in use and the created magnetic disturbance was recorded using commercial sensors (FieldLine V2). 

\begin{figure}[t]
\includegraphics[width=\columnwidth]{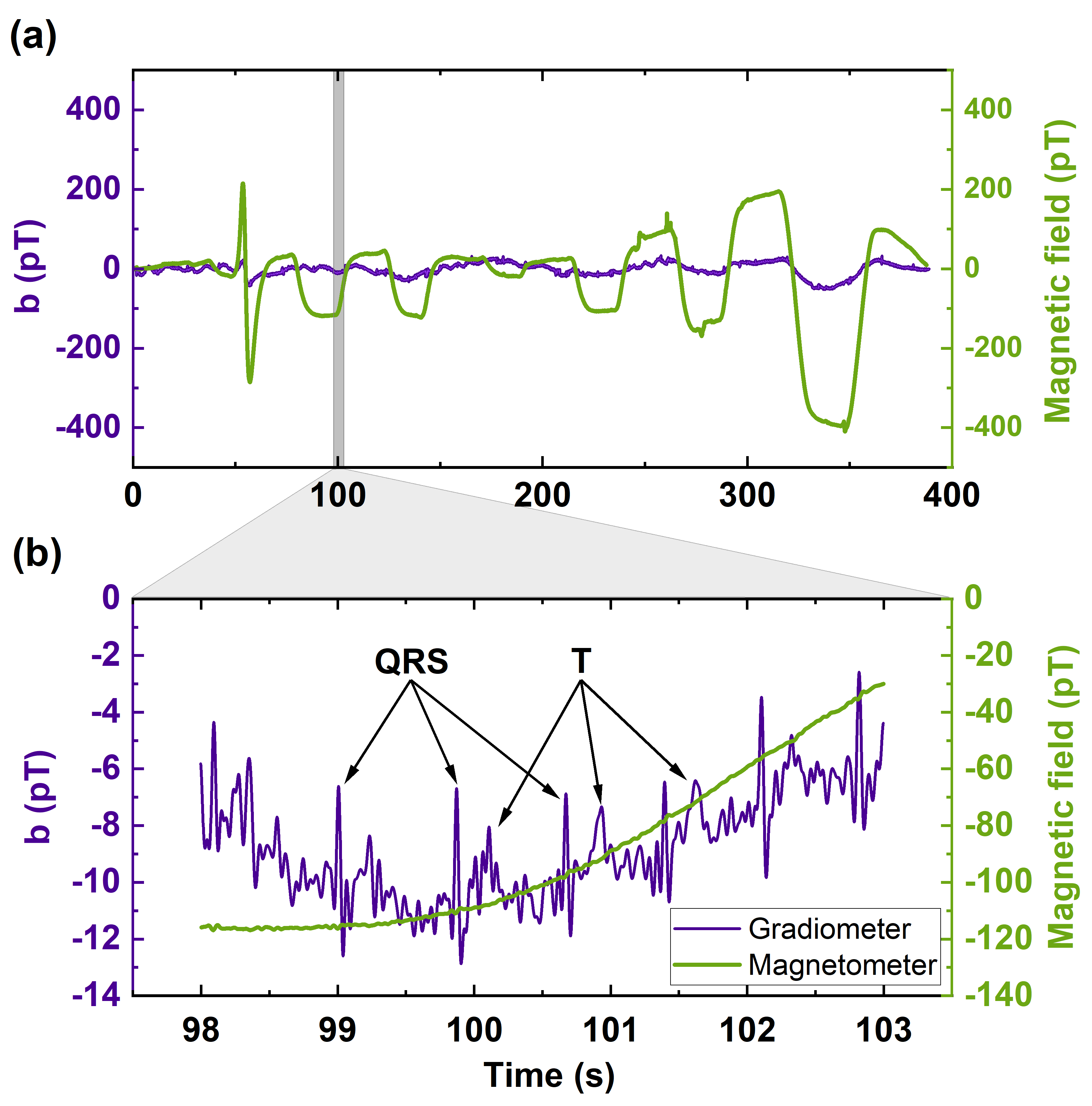}
\caption{\label{fig:heart} Magnetocardiogram recording in the gradiometer mode with the nearby lifts running:  (a) full duration of the recording, (b) zoomed-in area highlighted in grey in figure (a).  The purple trace is the gradiometer signal $b=B'd$ and the green trace is the reference magnetometer showing the magnetic field component in the gradiometer's sensitive axis direction. At 50 s, the impact of a magnetic lock while opening and closing door is visible. After 70~s, the operation of the building's lifts has started. The arrows indicate the QRS complex and T wave of the measured heartbeat. }
\end{figure}

Fig.~\ref{fig:heart} shows our magnetocardiography results.
The purple line in panel (a) is a 400~s recording with our gradiometer, while the green line is the background magnetic field fluctuations  measured with the reference sensor positioned in the same direction as the gradiometer. The peak at around 50~s is caused by opening and closing our laboratory entrance door which has a magnetic lock. This door is about 5~m from our magnetically shielded room. The rest of the peaks and dips are due to the lift movements between various floors. After 250~s the lifts traveled larger distances creating also substantial magnetic field gradients, that are therefore picked up by our sensor. Overall, we achieved attenuation of external magnetic field disturbances up to 27~dB. Fig. 5 b shows a zoom over the shaded area of figure 5(a) and shows the recorded heartbeats (purple line) together with the reference signals. Note the different vertical scales. There are three expected cardiography signals: the P wave, the T wave, and the QRS complex. The latter is responsible for a large spike in signal, which should be most clear. The P wave is a smaller waveform that precedes QRS complex, whilst the T wave is a similar, more powerful signal that occurs afterwards \cite{ECG}. The QRS complex is clearly visible, and for most trials so is the T wave. The P component is not so clear to notice. 

\section{Conclusion} \label{sec:conclusions}

In summary, we have discussed a scheme for the realisation of an OPMG based on NMOR, and shown that it is insensitive to external homogeneous magnetic field noise and sensitive to magnetic field gradients. We have detailed our methods to practically implement the scheme discussed and build up a compact magnetic gradiometer sensor. We have tested the performance of our sensor in controlled conditions, allowing us to measure its best sensitivity to magnetic field gradients and the optimal resilience to external magnetic fields. The amplitude of the magnetic noise range here investigated, $\delta$B~$<$~5~nT, is typical for light magnetically shielded rooms close to strong noise sources such as lifts or urban traffic. We have shown that our sensor can be adapted to work in different conditions, depending on the amplitude of the magnetic field noise present. We have demonstrated that our OPMG has sufficient sensitivity to measure human biomagnetism. In particular we have been able to record the auditory evoked response of a human brain with excellent signal-to-noise ratio, and we show that the measured brain response remains phase-locked to the stimulus over the duration of the recording. Finally we have demonstrated the ability to measure cardiac signals in real-time, even in the presence of significant transient magnetic field fluctuations. Our work provides new opportunities in measuring human biomagnetism, complementing the features of OPMs. The capabilities of atomic vapour sensors now match those of SQUIDS, but with enhanced performance, portability and reduced costs. Particularly interesting is the ability of combining our OPMG, whose head does not contain any magnetizable part, with transcranial magnetic stimulation. This could open new avenues in understanding brain connectivity and in development of drug-free treatments for various brain disorders.  

\section*{Acknowledgments} \label{sec:acknowledgments}

This work was supported by EPSRC (grant number EP/\allowbreak{}T001046/1). LK is supported by European Union's Horizon 2020 program\-me (No 101027633). OJ is supported by the Wellcome Trust Discovery Award (grant number 227420). AK is supported by EPSRC  Quantum Technology Career Development Fellowship (grant number EP/W028050/1).

\bibliography{Grad_bib}
\bibliographystyle{unsrt}
\end{document}